\documentclass[12pt,preprint]{aastex}
\def\lsim{\lower.5ex\hbox{$\; \buildrel < \over \sim \;$}}
\def\gsim{\lower.5ex\hbox{$\; \buildrel > \over \sim \;$}}

\begin{document}

\title{On the Ejection Mechanism of Bullets in SS 433}

\author{Sandip K.\ Chakrabarti$^{1,4}$, P.\ Goldoni$^{2}$, Paul J.\ Wiita$^3$,
A.\ Nandi$^{1}$, S.\ Das$^{1}$} 

\affil{$^1$S.N. Bose National Center for Basic Sciences, JD-Block, Salt Lake, Kolkata, 
700098\\ India\\
$^2$ Service d'Astrophysique, CEA/Saclay, 91191 Gif-sur-Yvette Cedex, France\\
$^3$ Department of Physics and Astronomy, Georgia State
University, Atlanta GA 30303\\
$^4$ also at Centre for Space Physics, P-61 Southend Gardens, 
Kolkata, 700084, India\\
e-mail: chakraba@bose.res.in, paolo@discovery.saclay.cea.fr, 
wiita@chara.gsu.edu, anuj@boson.bose.res.in, sbdas@bose.res.in}

\begin{abstract}
We discuss plausible mechanisms to produce bullet-like
ejecta from the precessing disk in the SS 433 system. We show that 
non-steady shocks in the sub-Keplerian accretion flow can provide the
basic timescale of the ejection interval
while the magnetic rubber-band effect of the toroidal flux 
tubes in this disk can yield flaring events.  
\end{abstract}


\noindent : To BE PUBLISHED IN APJL (SEPT 1st, 2002)

\section{Introduction}

SS 433 remains one of the most enigmatic objects 
in the sky. Even twenty-five years after its first
appearance in the catalogue of Stephanson \& Sanduleak
(1977), it is not clear whether the
compact object is a black hole or a neutron star. However,
there is ample evidence that the companion is an OB type star with
an orbital period of 13.1 days, which is
losing mass at the rate of about $10^{-4} M_\sun ~{\rm yr}^{-1}$ 
(van den Heuvel 1981), corresponding 
to extremely super-Eddington accretion regardless
of the mass of the compact object.  

One of the most curious properties of the jets of SS 433, which first made
their presence distinctly felt through the emission of variable H$\alpha$
lines, is that they are apparently ejected as bullets 
(e.g.\ Borisov \& Fabrika 1987;\ Vermeulen et al.\ 1993; Paragi et al.\ 1999; 2002;
Gies et al.\ 2002), with a surprisingly
nearly constant radial velocity of about $0.26c$.
The absence of a significant intrinsic rotational velocity (i.e., $v_\phi$)
component is clear from the fact that the kinematic model
(e.g.\ Abell and Margon 1979), which assumes only radial 
injection, quite accurately explains the time variation of the red- and blue-shifts
of the H$\alpha$ emission from the jets with a period of 162 days, which is
attributed to the precession of the accretion disk about the compact object.
The radial velocity is less than the maximum
allowed sound speed of $c/\sqrt{3}$ and thus hydrodynamic
acceleration could, in principle, explain it. Therefore
one may not require a magnetic or electrodynamic acceleration processes
(e.g.\ Belcher \& MacGregor 1976; Lovelace 1976). 
However, the rather good collimation (Margon 1984; Paragi et al.\ 1999) 
supports the hypothesis that a substantial degree of confinement 
produced by toroidal flux tubes may be present. 
Gies et al.\ (2002) showed that the ratios of the H$\alpha$ emission
equivalent widths from the approaching and receding jets as a
function of precessional phase only could be nicely fit if these
emission components are bullet-like. Indeed, the recent 
Chandra X-ray Observatory discovery of X-rays at a distance of about $10^{17}$cm
from the center may result from the collision of such bullets (Migliari,  Fender, \& Mendez  2002).

SS 433 poses another interesting problem: it was pointed out by Chakrabarti 
(1999) and Das \& Chakrabarti (1999) that significant outflows are produced only 
when the accretion rate is such that the X-ray source is in a low/hard state,
and all the observational indications in other micro-quasars
also suggest that the jets are indeed produced in low/hard states 
(Corbel et al.\ 2001; Klein-Wolt et al.\ 2002).
However, it is difficult to imagine how SS 433 manages to remain
in the low/hard state with $10^{-4}M_\sun ~{\rm yr}^{-1}$ of wind matter 
ejected from its companion. The answer to this quandary probably lies 
in the recent results of
Paragi et al.\ (1999) and Blundell et al.\ (2000),
whose high resolution radio maps show that there is a large region of
roughly $50 {\rm AU}$ in radius which is filled with enough gas and dust to obscure the 
accretion disk and the base of the jets. They also found an equatorial
outflow.  Gies et al.\ (2002) present additional evidence from
observations of the ``stationary'' H$\alpha$ and He I lines
for an extended ``disk wind''. So it is distinctly possible that 
most of the matter from the donor is rejected either by centrifugal force
(Chakrabarti 2002) or by radiation force far outside the central accretion disk, and thus 
the compact object receives only a few times the Eddington rate ($\dot M_{\rm Ed}$) of 
its companion's wind matter to accrete. 
This consideration finds further support from the fact that the
kinematic luminosity of the jet itself is around $10^{39}$ erg s$^{-1}$ (Margon 1984),
which corresponds to about one Eddington rate for a $10M_\odot$ compact object. 

In numerical simulations of supercritical winds by Eggum, Coroniti \& Katz (1985) 
designed to model SS 433, it was shown that only a fraction of a percent of 
the infalling matter
is ejected from a radiation pressure supported Keplerian disk,
which indicates that the accretion rate must be at least 
$100 \dot M_{\rm Ed}$ if the accretion takes place through a Keplerian disk.
On the other hand, numerical simulations of a
sub-Keplerian disk by Molteni, Lanzafame \& Chakrabarti (1994)
suggest that about $15-20$ percent of matter is ejected as an outflow,
indicating that the accretion rate onto the compact object in
SS 433 need be at most a few $\dot M_{\rm Ed}$.
Similar simulations with different parameters yield situations 
where no steady shocks can form, even though two saddle-type 
sonic points are present (Ryu, Chakrabarti, \& Molteni 1997,
hereafter RCM); under these conditions large scale shock oscillations 
produce intermittent outflows instead of continuous outflows. 
Since the compact object is a wind accretor, a low-angular momentum,
sub-Keplerian flow is the most likely description of the accretion flow.
Indeed, the presence of sub-Keplerian flows in several other high mass 
X-ray binaries has now been verified (Smith, Heindl \& Swank 2002).

In this {\it Letter}, we present  a few scenarios leading to
ejection of matter as bullets in SS 433. We discuss four possible
ways to create blobs of matter emerging from  the disk and conclude that 
periodic ejection of the blobs by the large scale oscillation of an accretion 
shock (something like a  piston) may be the fundamental production
mechanism of the ``normal'' bullets.  The irregularly observed 
rapid flaring (Vermeulen et al. 1993)
could be understood in terms of the catastrophic collapse of
toroidal magnetic flux tubes, very similar to what has been argued
to be occurring in GRS 1915+105 (Vadawale et al. 2001; Nandi et al.\ 2001).  
In the next section we discuss these processes and their suitability
or unsuitability for SS 433. In \S 3, we present concluding remarks.

\section{Mechanisms to produce bullet-like ejecta from accretion flows}

In both the works of Eggum et al.\ (1985)
and Molteni et al.\ (1994) continuous ejection was
reported when a radiation pressure dominated Keplerian disk, or a sub-Keplerian
disk capable of producing a steady shock, were considered. However,
in SS 433 the basic ejection is bullet-like and since the 
size of the X-ray emitting region is smaller than $l_x\sim 10^{12}~$cm 
within which the material in the jets is already accelerated to $v_{jet} \sim 0.26c$ 
(Watson et al.\ 1986; Stewart et al.\ 1987),
the bullets are not expected to be delayed by 
more than $l_x/v_{jet} \sim 100~$s. Indeed, recent Rossi X-Ray Timing
Experiment (RXTE) observations
of hard X-rays from SS 433 indicated  variability  
on time scales of 50--1000 s (Safi-Harb \& Kotani 2002), roughly corroborating
this picture. In fact, a simultaneous measurement of a flare at $2$ GHz in the radio
(Kotani \& Trushkin 2001) and in hard X-rays (Safi-Harb
\& Kotani 2002) indicated a strong anti-correlation of radio and X-ray
fluxes, similar to what is observed in GRS 1915+105 
(Mirabel \& Rodriguez 1994). Moreover,  the X-ray luminosity
is very low ($\sim 10^{36}$erg s$^{-1}$) and is believed to come from the base of the
jets (Watson et al.\ 1986). It is believed to have a thermal origin and
EXOSAT (Watson et al.\ 1986) and GINGA (Yuan et al.\ 1995) observations
were adequately fitted with a thermal bremsstrahlung model with $kT \gsim 30$ keV. 
The overall spectral shape suggests that the source has always been in a 
standard low/hard state and so far no quasi-thermal emission expected
from a `Keplerian disk' has been detected. From the interaction of the jet with the supernovae remnant W50,
the lower limit of kinematic luminosity is found to be at least $10^{39}$erg s$^{-1}$ 
(Biretta et al. 1983; Davidson \& McCray 1980). This means that the mean mass outflow rate is around
$10^{18}$g s$^{-1}$, and if most of it is  in the form of bullets
ejected at 50--1000 s intervals,
the mass accumulated in each bullet should be in the range of $10^{19}$--$10^{21}~$g.

The above data implies that the essential features that one must explain
when attempting to produce bullets out of the accretion disks are:
(a) the disk should be a sub-Keplerian flow; (b) the object 
(black hole or a neutron star) and its surroundings should be in a low/hard state;
(c) bullets should be ejected in 50--1000 s time-scales under normal
circumstances; (d) the mass of each bullet should be around $10^{19-21}$g;
and finally, (e) there should be occasional flaring with an
anti-correlation of radio and X-ray emission. We now discuss several scenarios and present
what we believe to be the most probable picture of what is
going on in SS 433. The four processes are schematically shown in Fig.\ 1(a-d).

\subsection{Cooling  of the jet-base by Comptonization and separation of blobs}

It was shown by several numerical simulations that significant 
outflows are produced from regions very close to the inner edge
of the accretion flow, possibly from the centrifugal 
pressure dominated region (Molteni et al.\ 1994; Molteni et al.\ 2001).
These jets are launched subsonically, but quickly  pass through
the inner sonic point to become supersonic. In the subsonic region
while the matter moves slowly, the density is high and
the optical depth could be large enough ($\tau >1$) to undergo Compton
cooling (Fig.\ 1a) {\it provided there is a 
Keplerian disk underneath to supply soft photons}. 
A part of the outflow, which was subsonic previously
becomes supersonic because of this rapid cooling 
and separates from the base of the jet.
This separation of blobs is expected to occur at the 
sonic surface $r_c$ which is $\sim 2-3 r_s$, where, $r_s$ 
is the size of the centrifugal barrier; see Chakrabarti (1999).

This possibility, though attractive, and in fact likely to be
a major mechanism for rapid state change in objects like
GRS 1915+105 (Chakrabarti \& Manickam 2000),
is untenable in SS 433 because the latter is a wind 
accretor: thus no significant Keplerian  disk is expected
in this system to supply the soft photons, and indeed none has been detected so far
(Watson et al.\ 1986; Yuan et al.\ 1995).

\subsection{Resonance oscillation of accretion shocks in the presence 
of bremsstrahlung cooling} 

Numerical simulations of accretion flow show that
in cases where the cooling time-scale nearly matches the infall 
time-scale, a shock forms, but it then starts oscillating and ejects matter
quasi-periodically (Langer, Chanmugam \& Shaviv 1983;
Molteni, Sponholz \& Chakrabarti 1996, hereafter MSC; see Fig. 1b). 
In order to have an oscillation period of around $50~$s,
the shock must be located at the large distance of
$r_{s,{\rm MSC}} \approx 6400 r_g$ for a black hole of
mass $M=10M_\sun$, where $r_g=2GM/c^2$. 
The mass of the post-shock region is computed by equating the
bremsstrahlung (which we assume to be the  major cooling mechanism) 
cooling time and the infall time in the post-shock region (MSC):
\begin{equation}
T_{\rm MSC} \simeq \frac {\cal E}{\dot {\cal E}} \simeq \frac {r_{s,{\rm MSC}}} {v_f} \simeq
\Bigl( \frac{R r_{s,{\rm MSC}}}{r_g}\Bigr)^{3/2} \frac{r_g}{c},
\end{equation}
where, ${\cal E}$ is the specific thermal energy, $v_f$ is the infall velocity
and $R=(\gamma+1)/(\gamma-1) \simeq 4-7$
(these limits are for a strong shock with $\gamma=5/3$ and $\gamma=4/3$, respectively) 
is the compression ratio at the shock. Assuming the gas density ($n$) and
temperature ($T$) scale as $n \sim r^{-3/2}$ and $T\sim r^{-1}$ respectively, 
the mass of the sub-Keplerian  region of $r<r_s$
turns out to be $7 \times 10^{19}$g (with $M=10M_\sun$, $\gamma=5/3$). 
This is indeed of the same order 
as the mass of the bullets observed in SS 433. However, one has to have both the
angular momentum and energy  of the injected material comparable to the 
marginally bound values determined by the central object
in order to achieve such an oscillation. On the other hand, if
the mass expulsion from the system takes place at the similar radius 
of $r_{ex} \sim 10^4 r_g=2\times 10^{9} M/M_\odot ~$cm due to the centrifugal force, 
the specific angular momentum of the flow is approximately 
$[r_{ex}/(2r_g)]^{1/2}r_g c \sim 70 r_g c$,
which is very large compared to the marginally bound value of $2 r_g c$. 
So it is unlikely that this  mechanism works in SS 433.

\subsection{Non-steady and non-linear shock oscillation}

A standing shock can form in a sub-Keplerian flow
only if there are two saddle type sonic points
and the Rankine--Hugoniot relation is satisfied at least at one point in between
these two sonic points.  However, Chakrabarti (1990) showed that
there is a large region of the parameter space where there are two saddle
type sonic points but the shock-conditions are not satisfied. 
Even an initially supersonic accretion (such as the wind from the companion)
can fall into this category.

What will happen to such a realistic flow, especially when the specific entropy at 
the  inner sonic point is greater than that at the outer sonic point? 
RCM discovered that a flow injected with these parameters exhibits
yet another type of shock-oscillation (Fig.\ 1c). Here the shock searches 
for a stable location and oscillates without finding it. In
the first half of the cycle, the shock recedes far away, the post-shock region fills
up, but the accretion is essentially completely blocked. 
In the second half of the 
cycle, the shock pushes the matter into the black hole, thereby evacuating the
post-shock region. In a realistic simulation, RCM find that
while the ratio of actually accreted matter to the amount
available from the companion, $R_{\rm ai} \equiv {\dot M}_{\rm acc}/{\dot M}_{\rm inj}$ 
would be around $0.2$ during
the first half-cycle, $R_{\rm ai} \sim 1.3$ in the second half-cycle. 
The outflow was also found to be very large.
The time scale of oscillation was found to be $T_{\rm RCM} \sim (4000-6000) r_g/c$ 
for a $r_s\simeq 20 r_g$ whose infall time is only about $T_{\rm MSC} \simeq 
(R r_s/r_g)^{3/2}(r_g/c) \sim (350-400) r_g/c$. 
Thus, this type of oscillation takes about a factor 
of $R_T=T_{\rm RCM}/T_{\rm MSC} \sim 15$ times longer than the resonance oscillation
discussed in \S 2.2.
For a $50~{\rm s}$ oscillation, the location of the shock should be obtained 
from $(r_{s,{\rm RCM}}/r_g)^{3/2} \simeq (1/R) (50~{\rm s}/R_T) (c/r_g) \sim 10^4 $,
which gives, $r_{s,{\rm RCM}} \sim 450 r_g$ for a $10M_\odot$ black hole,
a more physically reasonable value.
Even though the size of the oscillating region goes down by a factor of $10$ or so,
compared to that involved in the resonance oscillation,
the ejected mass need not go down (even for the same accretion rate 
as in the earlier case). This is because nearly all of the accretion flow is accumulated in
half the cycle ($\sim 25~$s in this case) before being ejected (see Fig.\ 2 of RCM).

Another advantage of this type of non-steady shock oscillation is that it is driven by centrifugal
force and not by thermal cooling. Hence the result is generally independent of the accretion rate.
Thus as long as the viscosity remains low, equivalent to having the
Shakura-Sunyaev (1973) parameter, $\alpha \leq \alpha_c \simeq 0.015$ 
(Chakrabarti 1990),
and $\dot M_{\rm inj}$ remains fairly constant,
this oscillation, once established, could be sustained indefinitely.

\subsection{Magnetic rubber-band effect} 

In the event of increase in magnetic activity of the disk, as could happen
for instance when the accretion disk bends towards the binary companion
during its precessional motion,  it is not unlikely that a strong magnetic field
will be first intercepted, and then advected, toward the inner edge 
of the disk.  In this case the field will preferentially become toroidal due to 
shear in the rotating flow. Then, as has already been pointed out (Chakrabarti \& 
D'Silva 1994; Nandi et al.\ 2001) the acceleration due to magnetic tension 
\begin{equation}
a_T = -\frac{B_\phi^2}{4 \pi r (\rho_e  + \rho_i) } \sim  -\frac{B_\phi^2}{4 \pi r \rho_e},
\end{equation}
would be the dominant force in the post-shock 
region of the sub-Keplerian flow (Fig.\ 1d). 
Here, $r$ is the major radius of the toroidal flux tube
and $\rho_i$ and $\rho_e$ are the densities of the medium internal 
and external to the flux tube respectively. The last step in Eqn.\ (2) is written because  
$\rho_i<<\rho_e$ for a strong flux tube. Since $B_\phi \propto 1/r$
and $\rho_e \propto r^{-3/2}$, we get
\begin{equation}
a_T \propto r^{-3/2},
\end{equation}
thus increasing rapidly as the tube comes closer to the black hole, and even
surpassing the magnetic buoyancy,
\begin{equation}
a_{MB}=\frac{1-X}{1+X}\Bigl[\frac{\lambda_{\rm Kep}^2-\lambda^2}{r^3}\Bigr] \simeq  
\frac{\lambda_{\rm Kep}^2-\lambda^2}{r^3},
\end{equation}
where $X=\rho_i/\rho_e \rightarrow 0$, and $\lambda_{\rm Kep}$ and $\lambda$ are 
respectively the
specific angular momentum of a Keplerian disk and the disk under consideration. 
the accelerations in Eqns.\ (3) and (4) do cross over, since when very 
close to a black hole, 
$\lambda \rightarrow \lambda_{\rm Kep}$ for a sub-Keplerian flow.

The effect of magnetic tension is dramatic, and the inner part of the disk is 
evacuated in the Alfv\'en time scale: $r/v_A \sim (r/a_T)^{1/2} \lsim 0.1$s, for 
a $10M_\sun$ black hole with a realistic Alfv\'en speed, $v_A \simeq 0.1 c$
(Nandi et al.\ 2001). The enhanced plasma ejection along the axis 
presumably causes sporadic magnetic flare events which would be observable 
as radio outbursts, at the same time reducing the X-ray emission from the disk which forms the 
base of the jet. Recently, such effects may have been seen
(Safi-Harb \& Kotani 2002) where simultaneous observations
of 2 GHz radio and 2--20 keV X-ray fluxes from SS 433 have been made, and a clear dip
in X-ray flux is seen at the same time a strong radio flare is
observed. It is worth noting that similar anti-correlated 
variations are common during flares in GRS $1915+105$ (Feroci et al.\ 1999; Naik et al.\ 2001)
and we suggest that the flares in SS 433 originate in the same way.

\section{Concluding remarks}

In this {\it Letter}, we have studied various competing processes
for the creation of bullets which move ballistically in the 
jet of SS 433. We showed that blobs may be separated by: (1) Comptonization;
(2) shock oscillations due to resonance; (3) oscillations due to inherent unsteady
accretion solutions; (4) intense magnetic tension of the toroidal flux tubes.
We reject the first possibility because it requires a large
Keplerian disk, which is unlikely.  We are unable to distinguish at this stage 
which type of shock oscillation is more capable of producing bullet formation 
in SS 433, but we prefer the third possibility due to its 
impulsive and generic nature and smaller involved region. We believe that the fourth possibility of 
the inner disk evacuation should produce flaring events, but will occur rather rarely,
perhaps only once in a single precession period, when the magnetic field
of the companion is preferentially tilted towards the accretion disk during precessional motion.
This fourth mechanism  gives rise to an anti-correlation between radio and 
X-rays, perhaps already observed in SS 433 (Safi-Harb \& Kotani 2002). 

SKC, AN, and SD acknowledge a grant from the Department of Science
and Technology, India, and  financial support
from CEA/Saclay where part of this work was performed. PJW is grateful
for hospitality at the Department of Astrophysical Sciences,
Princeton University, and for support from the Research Program Enhancement
program at Georgia State University.

{}

\vfil\eject
\begin{figure}
\plotone{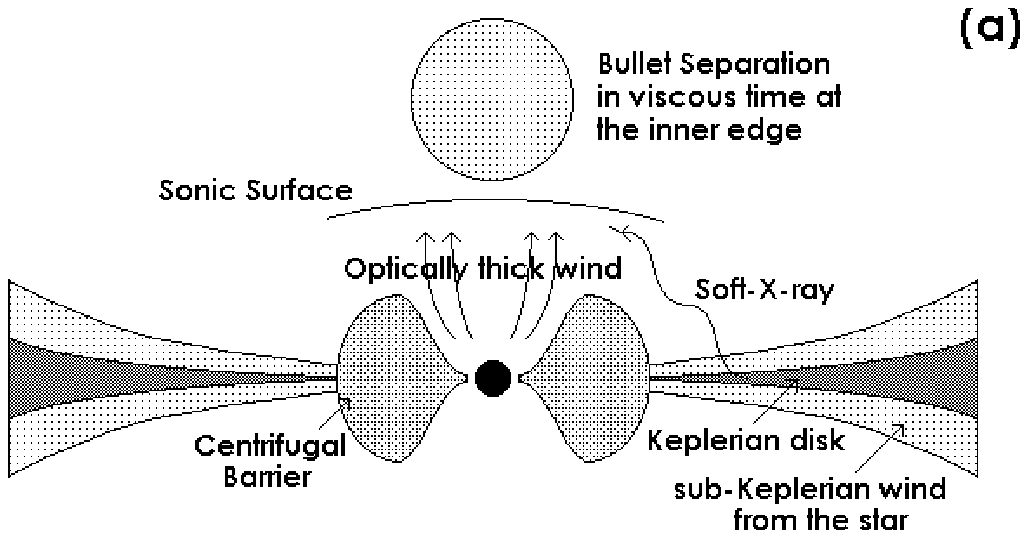}
\end{figure}

\vfil\eject
\begin{figure}
\plotone{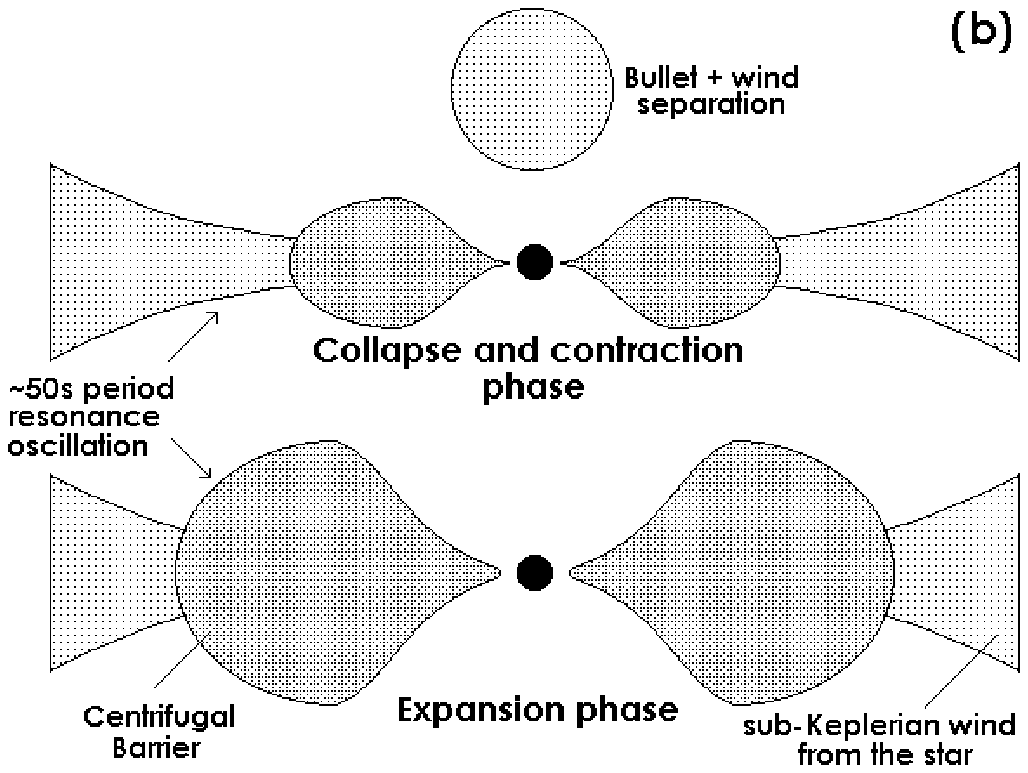}
\end{figure}

\vfil\eject
\begin{figure}
\plotone{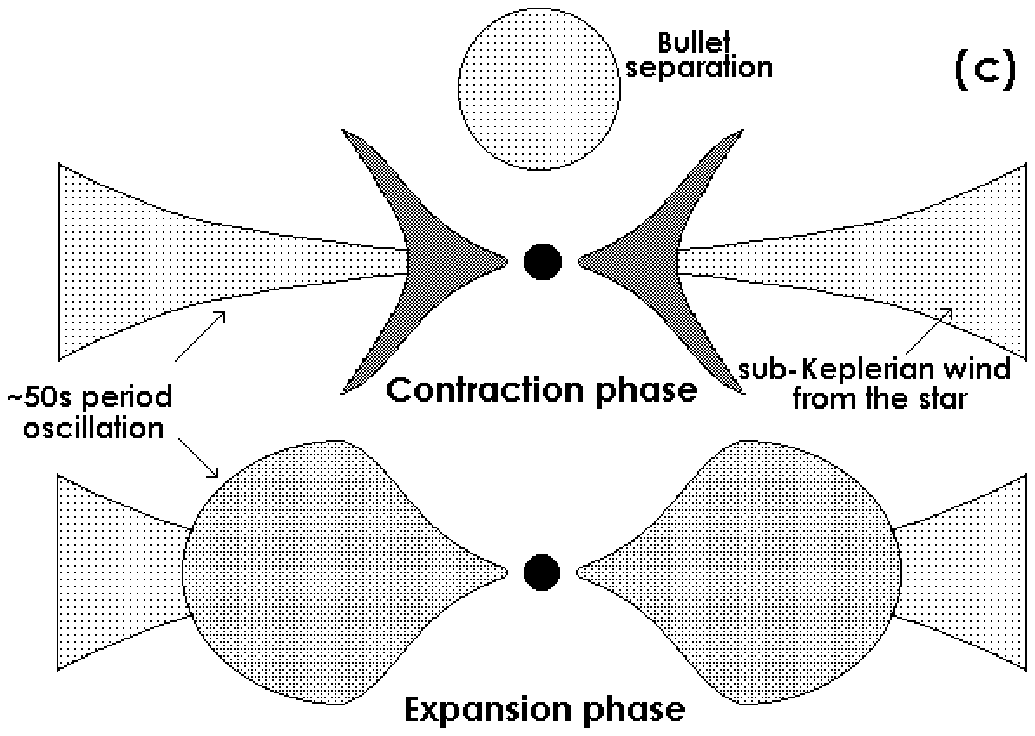}
\end{figure}

\vfil\eject
\begin{figure}
\plotone{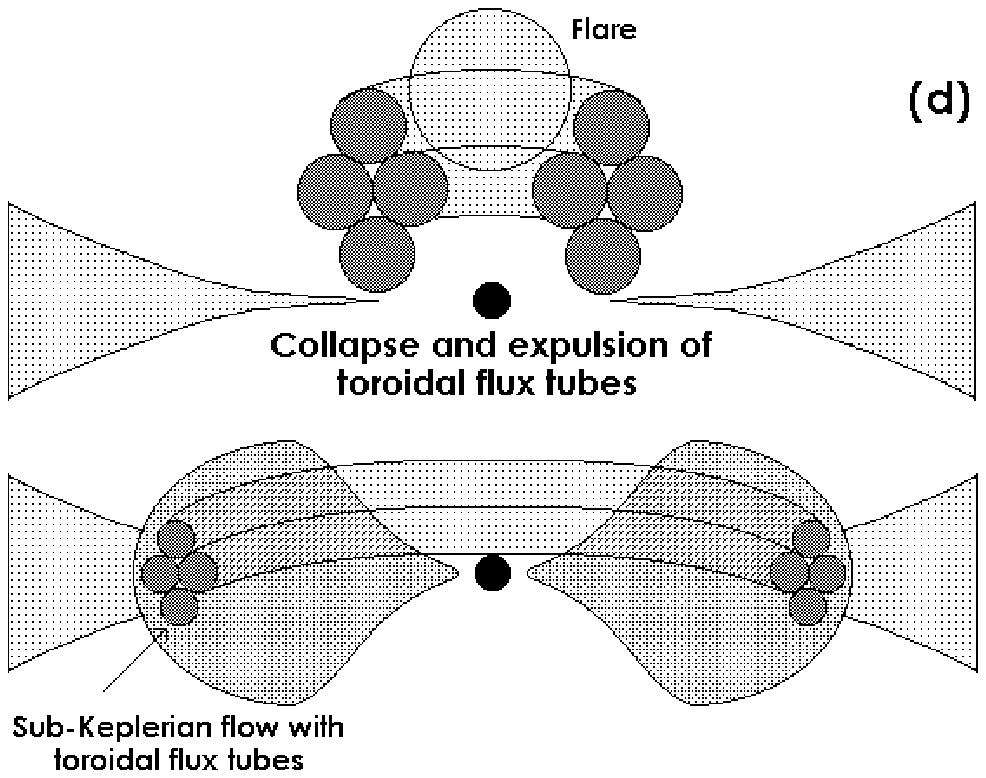}
\end{figure}

\begin{figure}
\caption {Four scenarios of bullet separation in SS433 are schematically 
shown. (a) The base of the jet is cooled down by soft photons from a Keplerian disk and 
detaches when it becomes supersonic.
(b) Resonance oscillation of the sub-Keplerian region due to the near matching of the
infall time with the cooling time produces discrete ejecta during the 
phase when the centrifugal barrier contracts. (c) Non-steady motion of the centrifugal 
barrier due to the inability of the flow to find a steady shock solution.  
(d) Magnetic tension from 
toroidal flux tubes (shown as shaded narrow tori) causes them to
collapse catastrophically in a hot ambient medium in rapid succession
which evacuates the centrifugal barrier. The recurrence time of: (a) is the viscous
time scale in the inner part of the disk, $\sim 10$s; (b--c) is $\sim 50$s; 
(d) is random and dictated by the enhanced magnetic activity. }
\end{figure}
 
\end{document}